\documentclass{aastex}

\begin{document}

\title{Dynamical Instabilities in Extrasolar Planetary Systems 
        Containing Two Giant Planets}

\author{Eric B.~Ford\altaffilmark{1}, Marketa Havlickova\altaffilmark{2}, and
        Frederic A.~Rasio\altaffilmark{3}}
\affil{Department of Physics, MIT, Cambridge, MA 02139}
\altaffiltext{1}{Present address: Department of Astrophysical Sciences, 
 Princeton University, Princeton, NJ 08544; eford@princeton.edu.}
\altaffiltext{2}{Also Department of Mathematics, MIT; mikihavl@mit.edu}
\altaffiltext{3}{Sloan Research Fellow; rasio@mit.edu.}

\slugcomment{submitted to Icarus, October 10, 2000}

\begin{abstract}
Instabilities and strong dynamical interactions between several giant
planets have been proposed as a possible explanation for the surprising
orbital properties of extrasolar planetary systems. In particular,
dynamical instabilities would seem to provide a natural mechanism
for producing the highly eccentric orbits seen in many systems.
Here we present results from a new set of numerical integrations for 
the dynamical
evolution of planetary systems containing two identical giant planets
in nearly circular orbits very close to the dynamical stability limit.
We determine the statistical properties of the
three main types of systems resulting from the development of 
an instability: systems containing one planet, following either a collision
between the two initial planets, or the ejection of one of them to infinity,
and systems containing two planets in a new, quasi-stable configuration.
We discuss the implications of our results for the formation and
evolution of observed extrasolar planetary systems. We conclude that the
distributions of eccentricities and semimajor axes for observed systems
cannot be explained easily by invoking dynamical interactions between
two planets initially on circular orbits. While highly eccentric orbits 
can be produced naturally
by these interactions, collisions between the two planets, which 
occur frequently in the range of observed semimajor axes, would
result in many more nearly circular orbits than in the observed sample.
\end{abstract}


\section{Introduction and Motivation}

The existence of planetary systems around other stars is now well
established.  For several years already, we have known many more planets
outside than inside our own Solar System (currently $\sim50$,
including planets around radio pulsars).  Several groups with ongoing
radial-velocity surveys have reported many unambiguous detections of
Jupiter-mass planets around nearby solar-like stars (for recent
reviews and updates, see Marcy and Butler 1998, 2000; 
Hatzes {\it et al.\/} 2000; Korzennik {\it
et al.\/} 2000; Perryman 2000; Vogt {\it et al.\/} 2000; Santos et
al.\ 2000).  Astrometry can sometimes help constrain the parameters of
the wider systems (e.g., Mazeh {\it et al.\/} 1999) and other, newer
techniques such as gravitational microlensing (Gaudi {\it et al.\/}
2000) and space interferometry (Fridlund 1999) are making rapid
progress.  Some of the most exciting recent developments include the
detections of planetary transits in HD 209458 (Charbonneau {\it et
al.\/} 2000; Henry {\it et al.\/} 2000) and two sub-Saturn-mass
candidates around HD 16141 and HD 46375 (Marcy {\it et al.\/} 2000).

\begin{figure}
\plotone{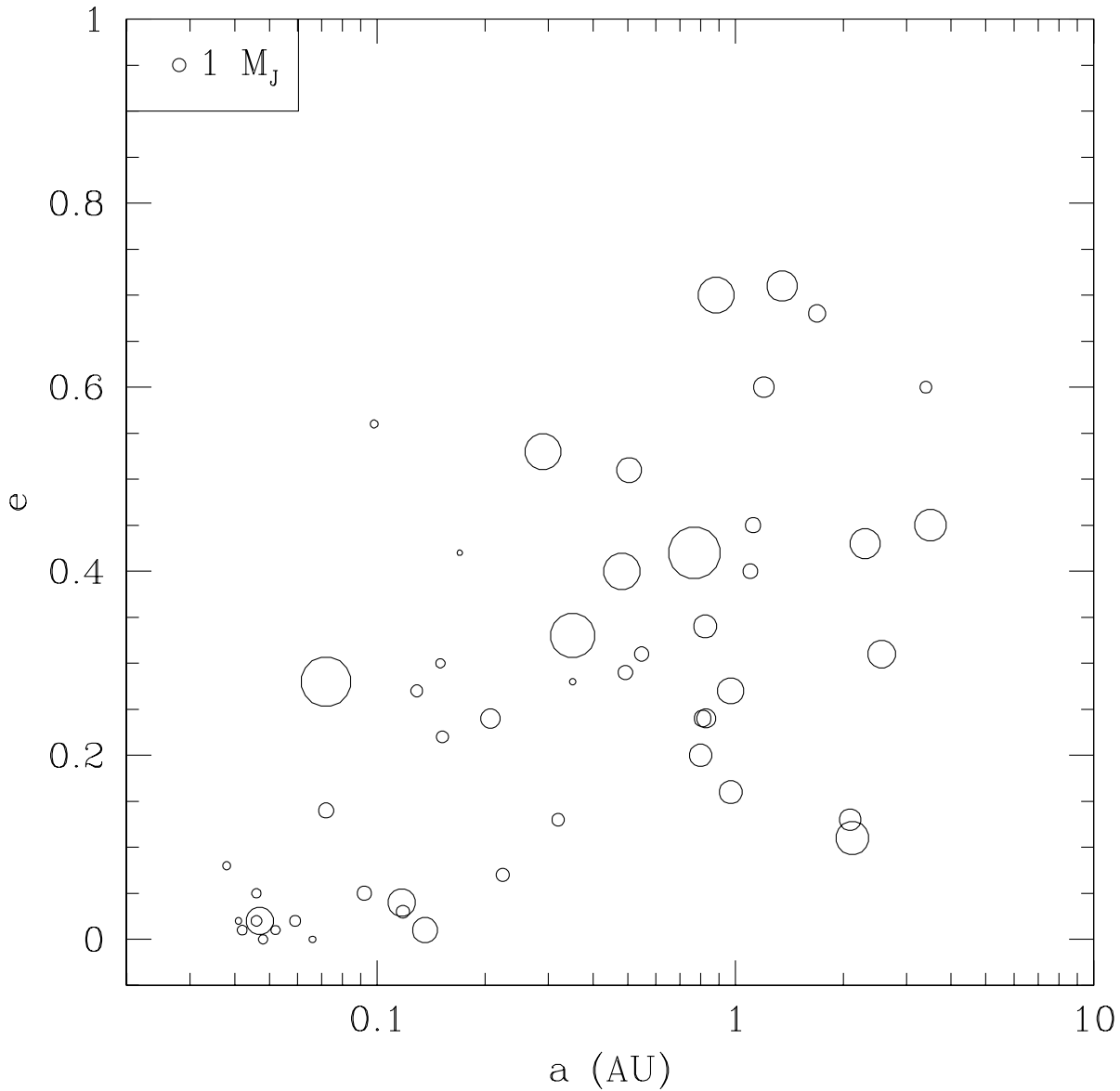}
\figcaption{Semimajor axes and eccentricities of all presently known
radial-velocity planetary candidates with $m \sin i< 13 M_{\rm J}$
($M_{\rm J}$ is the mass of Jupiter).
The area of each open dot is proportional to the value of $m \sin i$
for that object.  One group of planets has $a \la 0.07\,$AU and nearly
circular orbits, while the other, with $a \ga 0.07\,$AU, contains
many highly eccentric orbits. The data shown here include all planets
discovered up to and including the one in $\epsilon$ Eri (see Hatzes
{\it et al.\/} 2000 and references therein). Values of $m\sin i$,
$a$ and $e$ for 50 planets were taken from the table compiled by
Marcy {\it et al.\/} at {\tt http://exoplanets.org/} as of Oct 28, 2000.
 \label{FigEvsA}}
\end{figure}

The long-term stability of the Solar System, in spite of its chaotic
nature (see, e.g., Duncan and Quinn 1993), may have been necessary for
the development of intelligent life.  However, it may also be very
{\it atypical\/}, and may in fact require very special conditions
during the early stages of planet formation.  In particular, the
existence of a single dominant massive planet in our Solar System
(Jupiter), although perhaps essential for long-term dynamical
stability, may not be typical of planetary systems that form around
other stars. In fact, current theoretical models for the formation of
giant planets by accretion of gas from the nebula onto a rocky core,
combined with observed mass distributions of protoplanetary disks
(e.g., Beckwith and Sargent 1996) suggest that many planetary
systems could  form initially with 2--3 Jupiters. In a sufficiently
massive protoplanetary disk ($\ga 0.01M_\odot$), disk instabilities
may also lead naturally to the formation of several giant planets
(Armitage and Hansen 1999; Boss 1998). However, one should be careful
not to rely solely on theoretical models for planet formation that,
for many decades, have been based on only the Solar System for
guidance. What the newly detected systems (see Fig.~\ref{FigEvsA})
clearly tell us is that there exists a much greater variety of other
planetary systems than theorists had ever imagined. Some of the
unexpected properties of these systems (e.g., highly eccentric orbits)
suggest that, unlike our Solar System, they may well have been affected by
dynamical instabilities (Rasio and Ford 1996; Weidenschilling and 
Marzari 1996; Lin and Ida 1997).

In a system containing two or more Jupiter-like planets of comparable
masses the possibility exists that a dynamical instability will
develop, leading to strong gravitational interactions or collisions
between the planets (Gladman 1993, Chambers {\it et al.\/} 1996).
Here we will use numerical integrations of the orbital dynamics to
explore the consequences of such dynamical instabilities.  Based on a
preliminary set of calculations for systems containing two identical
giant planets (Rasio and Ford 1996) we expect a frequent
outcome of these instabilities to be a physical collision between two
giant planets. Very little mass is lost in such a collision, and the
result is therefore a more massive giant planet in a slightly more
eccentric orbit.  However, in many cases, the interaction can also
lead to the ejection of one planet to a larger distance while the
other is left in a slightly smaller, highly eccentric orbit.  If the
inner eccentric orbit has a short enough pericenter distance (distance
of closest approach to the star), it may later circularize through
tidal dissipation, leaving a Jupiter-type planet in a very tight
circular orbit around the star, with an orbital period typically of
order a few days.  Indeed, all the recently detected planets
around nearby stars are Jupiter-mass objects in very tight circular
orbits, or in wider eccentric orbits (Fig.~\ref{FigEvsA}).  The
standard model for planet formation in our Solar System (see, e.g.,
Lissauer 1993) is incapable of explaining them.  According to this
standard model, planetary orbits should be nearly circular, and giant
planets can only form at large distances ($\ga1\,$AU) from the central
star, where the temperature in the protostellar nebula is low enough
for icy materials to condense (Boss 1995, 1996). 

There are many ways for dynamical instabilities to develop in a
protoplanetary system.  The simplest scenario, that we will be
assuming here, is that two nearly identical giant planets had
initially formed (in the conventional way) at a large distance from
the central star, and later interacted dynamically (Rasio and Ford
1996). This could happen because their orbital radii evolved secularly
at different rates (significant orbital migration is thought to have
occurred in the outer Solar System; see Goldreich and Tremaine 1980,
Malhotra 1995), bringing them closer together, or because the masses
increased as the planets accreted their gaseous envelopes (Lissauer
1993), or both (see Kley 2000, who shows that a system of two
identical giant planets still embedded in a protoplanetary disk
generically evolves toward a dynamical instability).  The dynamical
instability leads eventually to orbit crossing and strong
gravitational interactions between the two planets (Gladman 1993).

Other formation mechanisms have been proposed for the 51-Peg-type
planets (in very tight circular orbits with orbital periods $\sim
3-5\,$d). If these planets had formed, like our own Jupiter, at a
large distance from the central star, some angular-momentum-loss
mechanism must have brought them in.  A {\it slow migration\/}
mechanism, such as friction in the protostellar nebula or interaction
with a protoplanetary disk, would tend to increase rapidly with
decreasing separation.  The dissipation would have had to switch off
at a critical moment for the planets to end up so close to the star
without being disrupted. Although there exist mechanisms that can
provide a barrier at some very short distance from the star (Lin {\it
et al.\/} 1996; Murray {\it et al.\/} 1998; Trilling {\it et al.\/}
1998), this always seems to require some fine tuning of the parameters
or extreme conditions such as a very massive disk (Murray {\it et
al.\/} 1998; Rasio {\it et al.\/} 1996).  Alternatively, direct {\it
in situ\/} formation of all 51-Peg-type planets by accretion onto a
solid core may also be possible under some conditions (Bodenheimer
{\it et al.\/} 2000).

Alternative mechanisms for inducing a large eccentricity in a planetary
orbit include the gravitational perturbation  by a distant
binary stellar companion (Holman {\it et al.} 1997; Mazeh {\it et al.\/} 1997)
and dynamical interactions with a gaseous protoplanetary disk 
(Papaloizou 2000) or with a disk of planetesimals (Murray {\it et al.\/} 1998).
Perturbations of planetary orbits in a binary star system are well 
understood theoretically (see, e.g., Ford {\it et al.\/} 2000;
Innanen {\it et al.\/} 1997)
and can lead to arbitrarily large eccentricities provided that the
relative inclination is large enough and the binary companion
is not too far from the planet.
While interactions with a disk typically {\it damp\/} orbital eccentricities,
they may in some special situations lead to modest eccentricity growth.
However, within our limited current theoretical understanding of these
processes (see, e.g., Nelson {\it et al.\/} 2000), it appears unlikely that they
would be able to produce eccentricities as large as those observed
for many Jupiter-mass objects (with measured eccentricities up to $e\simeq0.7$; 
see Fig.~\ref{FigEvsA}).

The strongest observational evidence to date that extrasolar planetary
systems may be affected by dynamical instabilities is provided by the
recent detection of two giant planets in wide eccentric orbits around
$\upsilon$~And (Butler {\it et al.\/} 1999). These are in addition to
the previously discovered 51-Peg-type planet, in a 4.6-day, nearly
circular orbit around the star. The presence of one or more additional
giant planets {\it in wider eccentric orbits\/} in the 51-Peg-type
systems is a basic theoretical prediction of our mechanism (see Rasio
and Ford 1996), since at least one other planet of comparable mass
must have been present to trigger an instability. Moreover, the two
outer orbits in $\upsilon$~And are tightly coupled (ratio of semimajor
axes $\simeq1/3$) and the system is still very close to the edge of
dynamical stability (numerical integrations indicate that it may in
fact be unstable on timescales $\sim 10^6-10^7\,$yr depending on the
precise values of the masses and orbital parameters; see Rivera and
Lissauer 2000).  This provides further evidence that the present
configuration resulted from the evolution of the progenitor planetary
system through a phase of violent dynamical instability.  Indeed,
systems of multiple planets that become unstable tend to evolve first
through a violent phase where energy and angular momentum are quickly
redistributed, followed by a much more gradual settling into a more
stable configuration that remains very close to the stability edge
(Chambers {\it et al.\/} 1996).

Additional support for a scenario based on dynamical instabilities comes
from the latest detection, around the nearby K2~V star $\epsilon$ Eri:
a planet with mass $m\sin i = 0.86\,M_{\rm J}$,
a long orbital period $P = 6.9\,$yr ($a\simeq 3.3\,$AU), and a 
large eccentricity $e\simeq 0.6$ (Hatzes {\it et al.\/} 2000). The star
is not in a binary system and, at this large orbital separation, the
planet is unlikely to have had a significant interaction with a
protoplanetary disk (which would also have produced significant
inward migration). This clearly leaves dynamical interaction with
another giant planet (which was likely ejected from the system)
as the most natural explanation.

While our study will concentrate on the systematic study of a system
containing two giant planets, other groups have performed small
numbers of exploratory calculations to determine the consequences of
dynamical instabilities in systems containing three planets or more.
Weidenschilling and Marzari (1996) have published results for the case
of three planets, while Lin and Ida (1997) presented results for
systems containing up to 9 planets. With many planets, successive
collisions and mergers can lead to the formation of a fairly massive
($\ga 10\,M_{\rm J}$) object in a wide, eccentric orbit.  We feel that
it is important to first understand fully the case of two planets.
One important advantage of the two-planet case is that the dynamical
stability boundary is very sharply defined, and its location known
analytically (Gladman 1993). Therefore, the initial value for the
ratio of semimajor axes $a_2/a_1$ must be varied only in a very narrow
range (right around the stability boundary) for each case. In
contrast, for three or more planets, the stability boundary is not
well defined (Chambers {\it et al.\/} 1996), and a much wider range of
semimajor axis ratios would have to be explored. In addition, if the
instability is triggered by the increase in the mass of one planet as
it accretes its gaseous envelope, we would expect that it would
naturally tend to involve only two giant planets, since it appears
extremely unlikely that more than two planets would be going through
the accretion process at precisely the same time.  Indeed a time
difference $\ga 10^6\,$yr between the formation of the different giant
planets would be expected from the standard scenario (Lissauer 1993).

\section{Methods and Assumptions}

\label{InitCond}

Our numerical integrations were performed for a system containing two
identical planets, with a mass ratio $m/M= 10^{-3}$, where $m$ is the
planetary mass and $M$ is the mass of the central star (This
corresponds to $m\simeq1\,M_ {\rm J}$ for $M=1\,M_\odot$). For this
system, the dynamical stability limit (for circular, coplanar initial
orbits) corresponds to $\alpha\equiv a_1/a_2 = 0.769$, where $a_1$
and $a_2$ are the semimajor axes of the two planets (Gladman 1993).
Our simulations were started with $\alpha$ randomly chosen in the
range from 0.769 to 0.781 (See Sec.~3 for a justification of this
range). The initial eccentricities were distributed
uniformly in the range from 0 to 0.01, and the initial relative
inclination in the range from 0 to $5^{\circ}$. All remaining angles
(longitudes and phases) were randomly chosen between 0 and $2\pi$.
Throughout this paper we quote numerical results in units such that
$G=a_1=M=1$. In these units, the initial orbital period of the inner
planet is $P_1\simeq 2\pi $.

The orbital integrations were performed using a modified version of
SWIFT, a standard software package for orbital dynamics developed by
Levison and Duncan (1994).  The package features several integrators,
including a Bulirsch-Stoer (BS) integrator and a mixed variable
symplectic integrator (MVS).

The BS integrator directly solves the second-order differential
equations of motion. For this work we have modified the BS integrator
to allow for regularization: whenever the outermost planet is
sufficiently distant from the inner planet(s) and the central star,
the outer planet is analytically advanced in its orbit and the motion
of the inner planets is integrated separately for the time necessary
to bring the outer planet back within the specified distance of other
planets.  The direct integration of the dynamical equations then
continues according to the original BS integrator. For the case of two
planets, this leads to a phase of the evolution where both planets are
following unperturbed Keplerian orbits. In our specific case, whenever
$r_2/r_{{\rm apo},1} > 100$, we switch to analytic Keplerian orbits.

The MVS integrator exactly solves an approximation of the system's
Hamiltonian (Wisdom and Holman 1991, 1992).  While the MVS integrator
is nearly an order of magnitude faster, it cannot handle close
encounters between the planets. Therefore we only use the MVS
integrator to determine the location of the stability boundary, and to
integrate all systems up to the first strong interaction.  Then the
regularized BS integrator takes over to follow the evolution of
unstable systems.

Throughout the integrations, close encounters between any two bodies
were logged, allowing us to present results for any values of the
planetary radii using a single set of orbital integrations. However,
the integration was stopped if the two planets collided with an
assumed minimum radius $R_{\rm min}/a_1=0.1\,R_{\rm J}/5\,{\rm AU}=
0.95\times10^{-5}$ (where $R_{\rm J}=7\times10^9\,$cm is Jupiter's
radius), or if either planet came to within a distance $r_{\rm
min}/a_1= 10\,R_\odot/1\,{\rm AU}= 0.01$ of the star (see below).

The BS integrations were performed using an accuracy parameter of
$10^{-12}$, which is used to determine each stepsize.  All integrations
conserved total energy and angular momentum within $10^{-4}$, although
for most runs energy and angular momentum were conserved to $10^{-6}$.
The computations were performed on the SGI/Cray Origin2000
supercomputers at the National Center for Supercomputing Applications
and at Boston University, and on the Condor cluster of Sun
workstations operated by the University of Wisconsin.  
The results
presented in this paper are based on $\sim10^3$ numerical integrations all
performed for systems with initial parameters in the ranges specified
above\footnote{The initial parameters used in the preliminary study
by Rasio and Ford (1996) were slightly different, and the shorter
numerical integrations did not provide a good basis for a statistical
analysis of the final properties of unstable systems.}.  
Each run was terminated when one of the following four
conditions was encountered: (i) one of the two planets became unbound
(which we defined as having positive energy, a positive radial
velocity, and being at least 100 times as distant from the star as the
other planet); (ii) a collision between the two planets occurred
assuming $R=R_{\rm min}$; (iii) a close encounter occurred between a
planet and the star (defined by having a planet come within $r_{\rm
min}$ of the star); (iv) the integration time reached $t_{\rm max}=
10^7$ (corresponding to about $1.6\times10^6\,P_1$).  The percentages of
runs that terminated according to each condition were approximately: (i) 50\%; (ii)
5\%; (iii) $<1$\%; (iv) 45\%. The total CPU time required for this study was
about 12,000 hours, corresponding to an average of about 12 CPU hours
per run.

Since collisions of a planet with the star seem to occur so rarely, only three
types of outcomes will be discussed in the rest of the paper. These three
types will be referred to as ``collisions,'' meaning a collision
between the two planets, ``ejections,'' meaning that one planet was ejected to
infinity, and ``two planets,'' meaning that two bound planets remained in a
(possibly new) dynamically stable configuration.

\section{Results}

\begin{figure}
\scalebox{0.95}{
\plotone{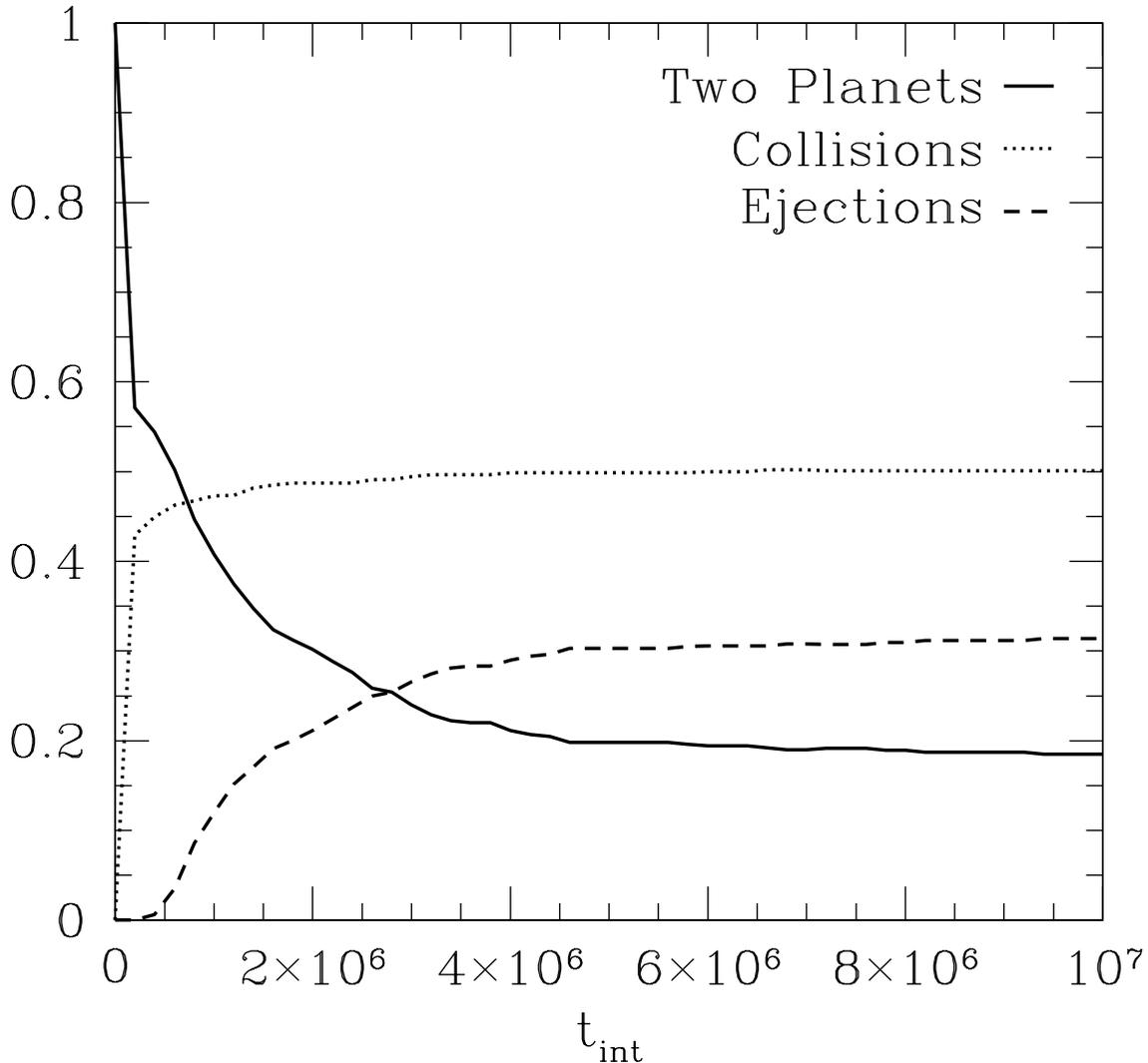}
}
\vspace{-5mm}
\figcaption{Branching ratios of various outcomes as a function of maximum
integration time (here and throughout this paper units are defined by
$G=a_1=M=1$, where $a_1$ is the initial semimajor axis of the inner
planet and $M$ is the mass of the central star).  The dashed line
corresponds to the ejection of one planet from the system, the dotted
line to a collision between the two planets (here assuming that the
planetary radius is given by $R/a_1 = 3\,R_{\rm J}/5\,{\rm
AU}=2.9\times10^{-4}$, and the solid line to cases where both planets
remain in a bound configuration. The branching ratios are well
determined for $t_{\rm int}=t_{\rm max}=10^7$, used in the rest of
this paper.\label{FigTime}}
\end{figure}

\begin{figure}
\plotone{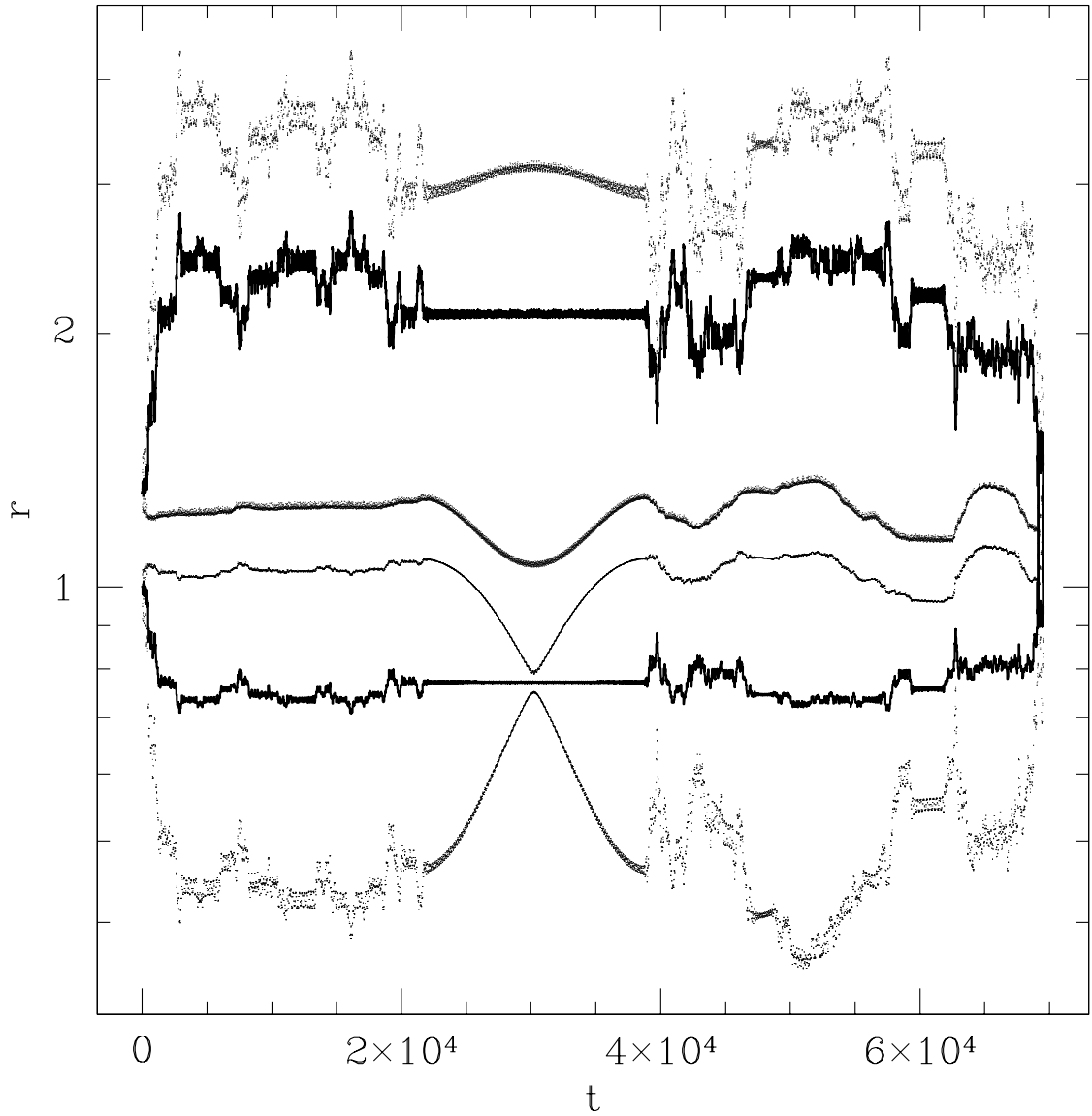}
\figcaption{Typical evolution of a system resulting in a collision between the 
two planets (at far right).  The two solid lines show the osculating semimajor 
axes of the two planets.  The dotted lines show the osculating pericenter 
and apocenter distances for each of the two planets. \label{FigExCollide} }
\end{figure}

\begin{figure}
\plotone{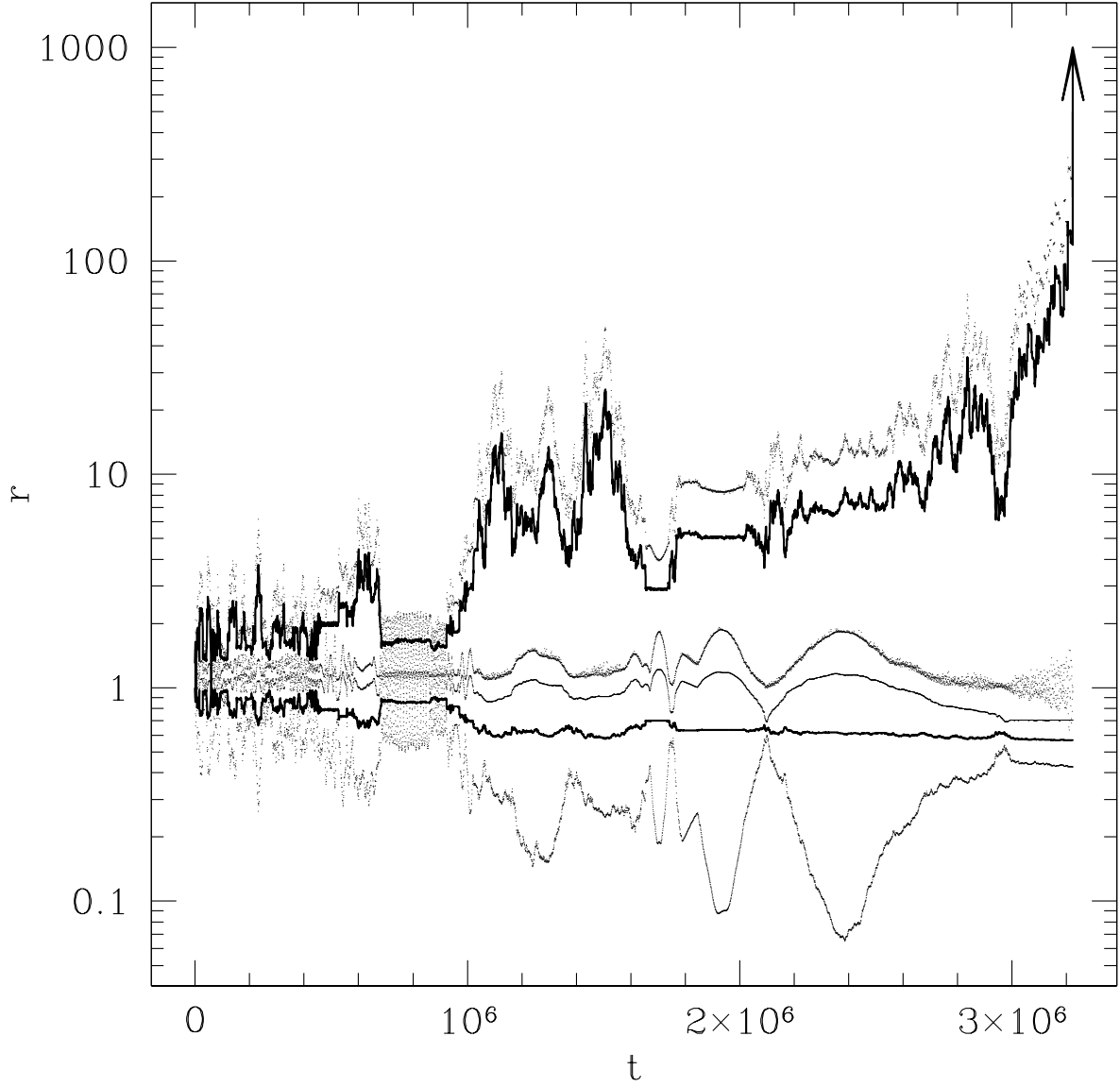}
\figcaption{Typical evolution of a system resulting in one planet being 
ejected to infinity (at arrow).  Conventions are as in Fig.~\ref{FigExCollide}. 
\label{FigExEject}}
\end{figure}

\begin{figure}
\plotone{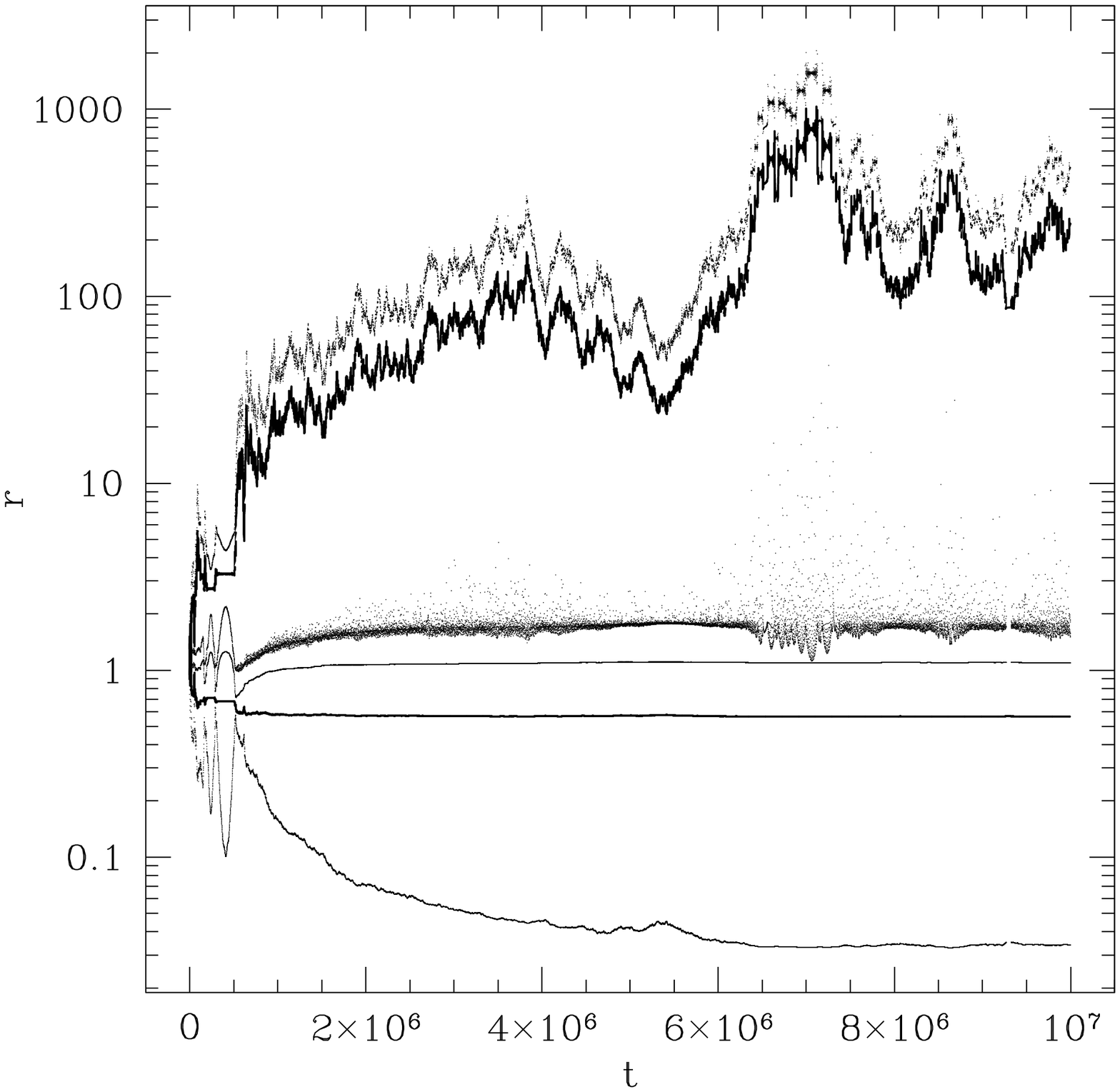}
\figcaption{Typical evolution of a system that retains both planets 
following a period of strong dynamical perturbations. Here the two planets are
still on bound orbits at the end of the integration, $t=t_{\rm max} = 10^{7}$.  
Conventions are as in Fig.~\ref{FigExCollide}.\label{FigExUnres}}
\end{figure}

For a study of this type to be meaningful, it is crucial to establish
that the integration time is long enough for the system to have
reached its true final configuration.  Fig.~\ref{FigTime} shows the
branching ratios for the three types of outcomes as a function of
integration time, for an assumed planetary radius $R/a_1 = 3\,R_{\rm
J}/5\,{\rm AU}= 2.9\times10^{-4}$.  While collisions nearly always
occur soon after the development of the instability (within $t\sim
10^5$, nearly independent of the planetary radius, see
Fig.~\ref{FigExCollide}), ejections can take a much longer time. This
is because the exchange of energy between the two planets typically
takes place through a large number of very weak interactions, rather
than just one strong interaction\footnote{For this reason also, the simple
arguments presented by Katz (1997), assuming a single strong interaction
between two planets, are irrelevant for real systems.} 
(see Fig.~\ref{FigExEject}).  We see
in Fig.~\ref{FigTime} that for integration times $t_{\rm int}\ga
6\times10^6$, the branching ratios become nearly constant, as desired.
All results shown in the rest of this section correspond to an
integration time $t_{\rm max}=10^7$.  
Fig.~\ref{FigExUnres} shows the evolution of a typical
system for which our numerical integration was terminated at 
$t=t_{\rm max}$ while the two planets were still
in a bound configuration, even though a strong dynamical instability
had clearly developed.

\begin{figure}
\plotone{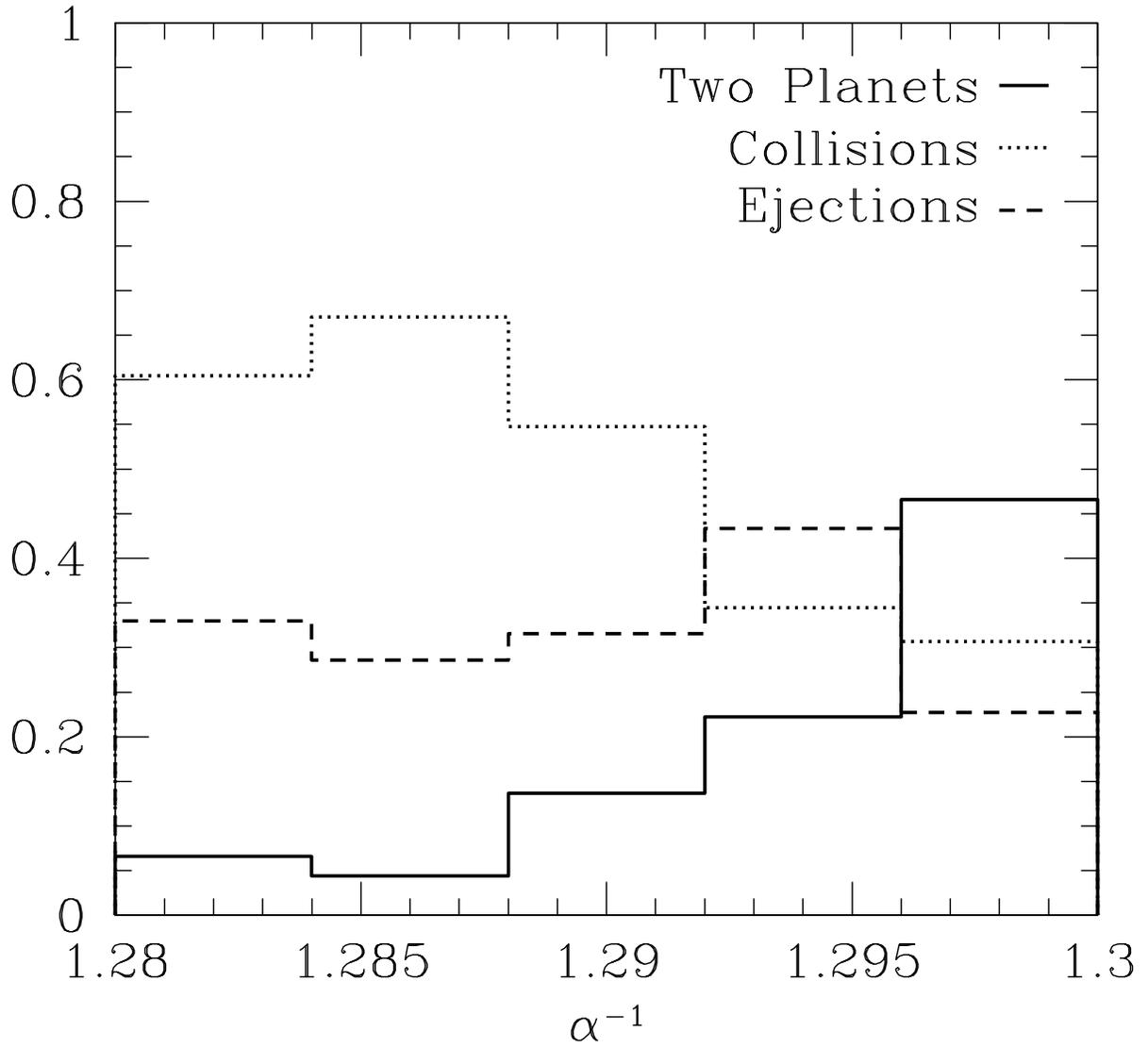} 
\figcaption{Branching ratios of different outcomes measured for different
ranges of values of $\alpha^{-1}=a_2/a_1$ (the initial ratio
of semimajor axes). Conventions are as in Fig.~\ref{FigTime}. 
See text for discussion.
\label{FigSemi} }
\end{figure}

We have also checked that the distributions of orbital parameters
determined for each type of final outcome are independent of the
precise values of the initial parameters used within the narrow ranges
considered. For example, we see no statistically significant variation
in the distributions of final orbital parameters measured for systems
starting in different sub-ranges of values for $\alpha$.  This
suggests that the properties of systems affected by an instability
will be largely independent of the particular mechanism triggering the
instability (which may determine the exact location of the system near
the stability boundary).  Similarly, we have verified that our results
are independent of the precise range of small eccentricities and
inclinations assumed in constructing the initial conditions.  However,
we find that, as expected, the {\it branching ratios\/} for different
types of outcomes do show a significant dependence on $\alpha$
(Fig.~\ref{FigSemi}). Indeed, for systems very near the edge of
stability, we expect that the branching ratio for retaining two
planets in a stable configuration should approach unity, while it
should go to zero further away into the unstable region. From
Fig.~\ref{FigSemi} we see that the transition region extends from the
theoretical stability edge at $\alpha^{-1}=a_2/a_1=1.3$ (all systems
with $a_2/a_1>1.3$ must be stable; see Gladman 1993) down to
$\alpha^{-1}=a_2/a_1=1.28$, where the probability of retaining two
planets in a stable configuration goes to nearly zero. Systems
entering the unstable region {\it slowly\/} (i.e., on a timescale long
compared to the typical growth time of dynamical instabilities,
$t_{\rm dyn}\sim 10^4-10^5\,$yr) will populate the entire range of
initial values of $\alpha$ shown in Fig.~\ref{FigSemi} (justifying our
choice of this range).  Systems entering the unstable region more
rapidly may "overshoot" our range of initial values for $\alpha$. To
model such a rapid evolution correctly would require the inclusion of
additional forces (e.g., from hydrodynamics) and is beyond the scope
of this paper.

\begin{figure}
\plotone{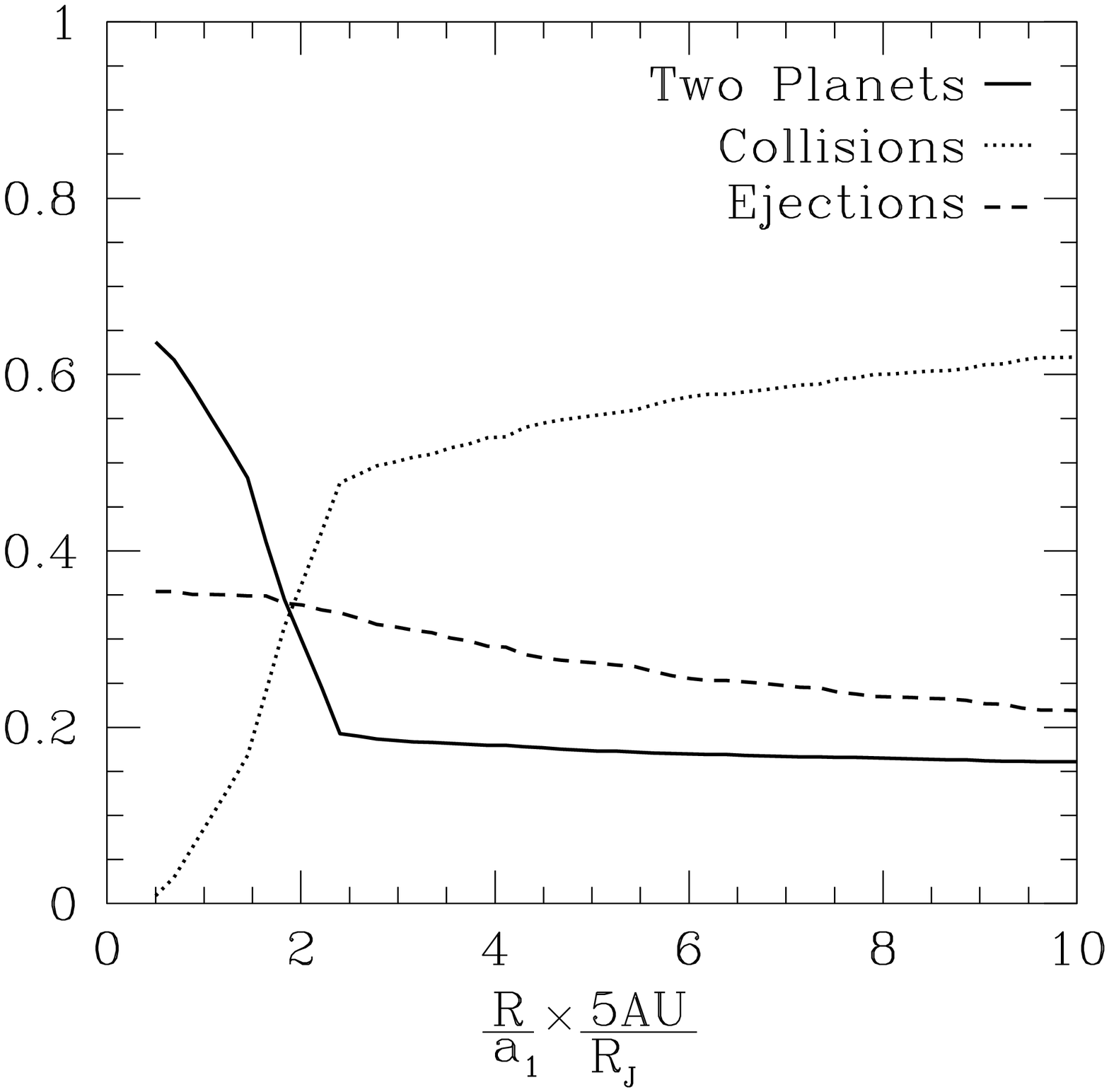}
\figcaption{Branching ratios for various outcomes as a function of
planetary radius $R$. Conventions are as in Fig.~\ref{FigTime}. Note how
the most probable outcome makes a sharp transition to favor collisions
as the planetary
radius increases or the semimajor axis decreases. \label{FigRadius}}
\end{figure}

The initial values of the planetary radii can also affect
significantly the outcome of a dynamical instability
(Fig.~\ref{FigRadius}).  We clearly expect larger planets to collide
more often, but Fig.~\ref{FigRadius} reveals that the fraction ejected
is only slightly reduced as the planetary radius increases.
Instead, as the 
radius increases, the branching ratio for collisions increases mainly
at the expense of the branching ratio for retaining two planets. This
is again a consequence of the mechanism for ejections, which proceed
through a large number of distant, weak encounters between the two
planets.  Since the radius of a giant planet depends extremely weakly
on its mass, most of the variation in $R/a_1$ for different planetary
systems will come from the initial semimajor axis $a_1$. For
$R=1\,R_{\rm J}$, collisions will be very rare in a system with
$a_1=5\,$AU, but they will occur for over 40\% of unstable systems
with $a_1=1\,$AU. Note that the radius of a newly formed giant planet
can be significantly larger than its radius today (see, e.g., Burrows
{\it et al.\/} 1997). For Jupiter, the initial radius may have been as
large as $\sim 2\,R_{\rm J}$, implying that even for a system with
$a_1=5\,$AU collisions may be significant.

\label{CollideText}

\begin{figure}
\epsscale{0.70}
\plotone{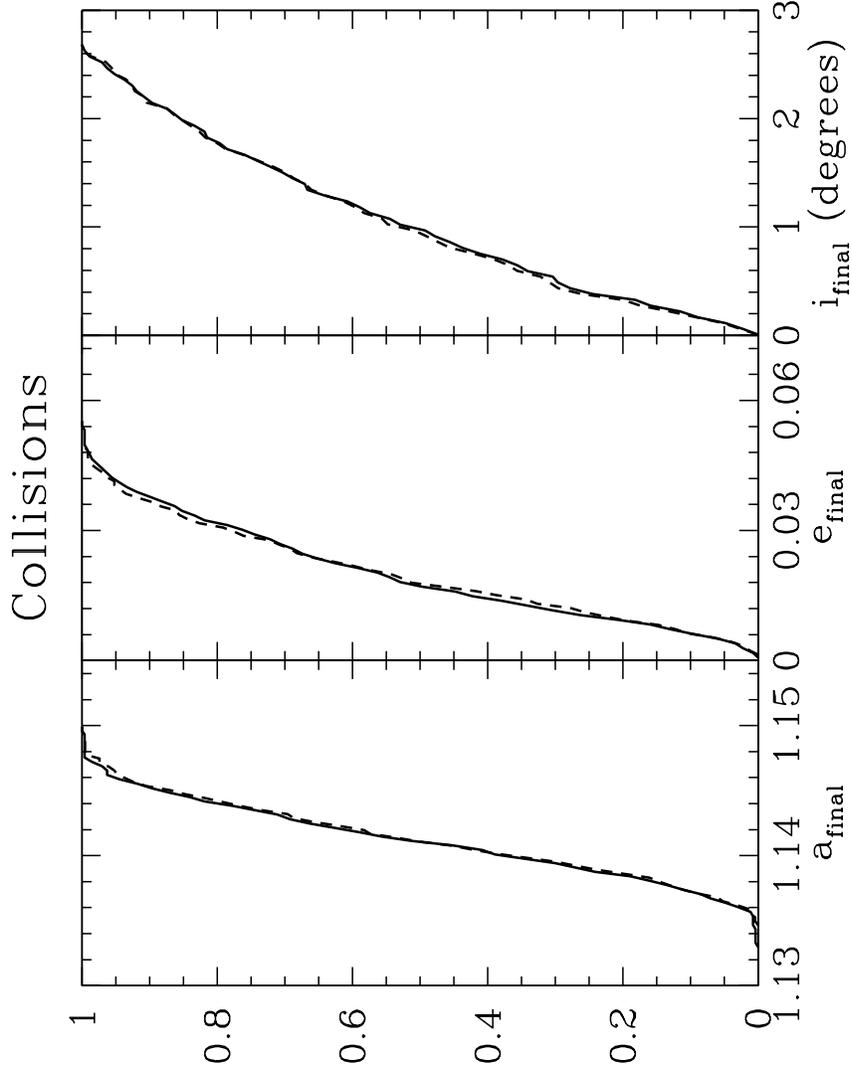} 
\figcaption{Cumulative distributions of the final semimajor axis (left),
eccentricity (center), and inclination (right; measured with respect
to the initial orbital plane of the inner planet) of the single planet
remaining after a collision (assumed to conserve mass and
momentum). The solid and dashed lines are for different planetary
radii, with $(R/a_1)\times(5\,{\rm AU}/R_{\rm J}) = 5$ and~3,
respectively. All planets resulting from a collision have very small
eccentricities ($e_{\rm final} \la 0.05$) and very small inclinations
(less than a few degrees).  The final semimajor axis is intermediate
between the two initial semimajor axes (recall that $a_1=1$ and
$a_2\simeq 1.3$ in our units). \label{FigCollide}}
\end{figure}

Collisions leave a single, larger planet in orbit around the star.
The energy in the center-of-mass frame of the two planets is always
much smaller than the binding energy of a giant planet (see Appendix).
As a consequence, collisions between giant
planets resemble parabolic collisions between low-mass main-sequence
stars in globular clusters. Hydrodynamic calculations show that, as
expected from simple energetic arguments, these collisions produce
very little mass loss (typically $\la5\%$; see, e.g., Lombardi {\it et
al.\/} 1996).  Therefore, to a very good approximation, we can model
the collisions as completely inelastic and assume that the two giant
planets simply merge together while conserving total momentum and
mass. Under this assumption, we have calculated the distributions of
orbital parameters for the collision products (Fig.~\ref{FigCollide}).
The final semimajor axis lies typically just inside the average of the
two initial semimajor axes. Eccentricities and inclinations remain
very small. These results are consistent with expectations from
elementary analytic arguments based on our assumptions
(see Appendix).

\label{EjectText}

\begin{figure}
\epsscale{0.80}
\plotone{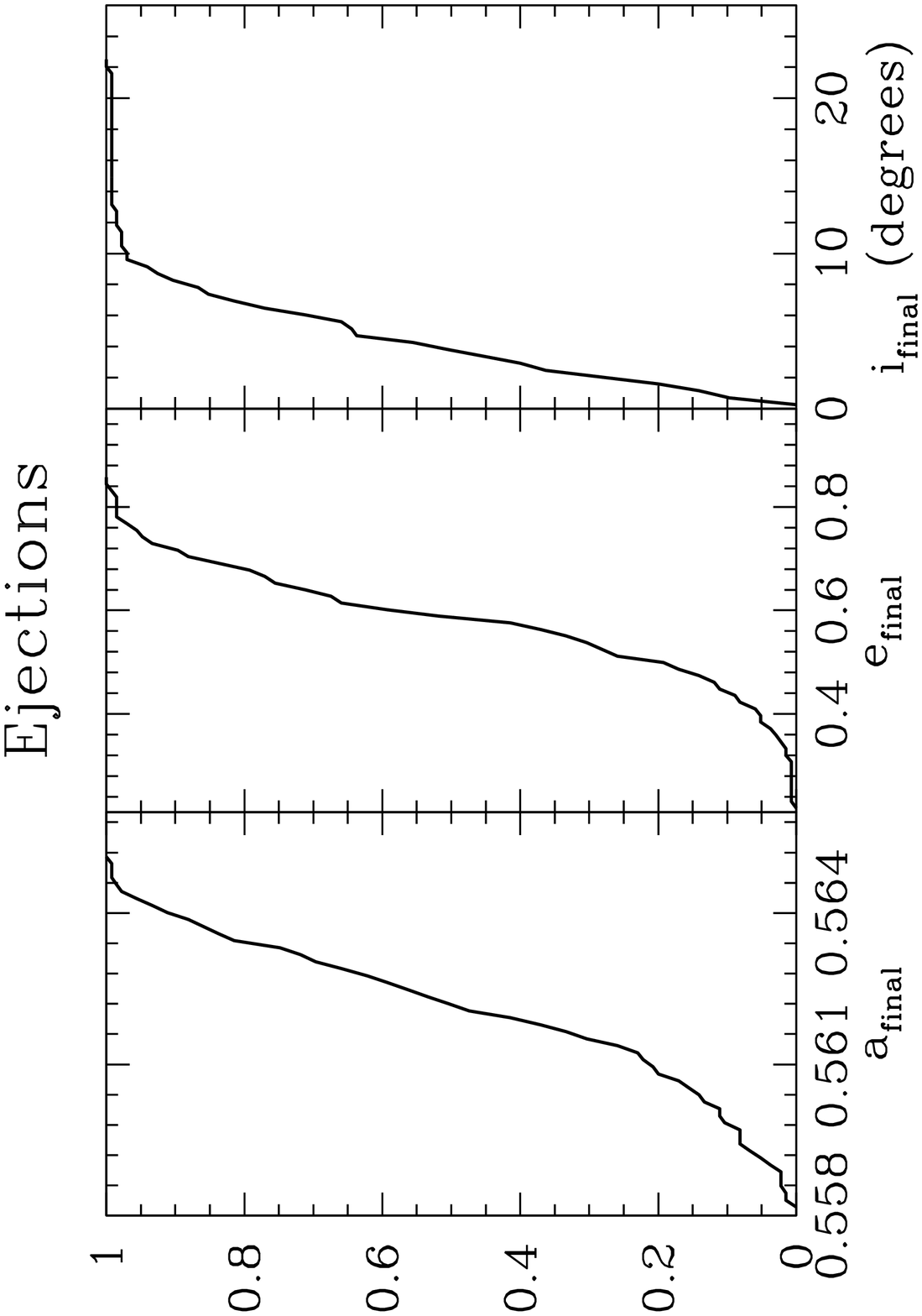} 
\figcaption{Cumulative distributions of the final semimajor axis,
eccentricity, and inclination of the remaining planet following
ejection of the other. Since the
number of ejections is almost independent of planetary radius (see
Fig.~\ref{FigRadius}), we only show results for $(R/a_1)\times(5\,{\rm
AU}/R_{\rm J}) = 5$. \label{FigEject} }
\end{figure}

\begin{figure}
\epsscale{0.80}
\plotone{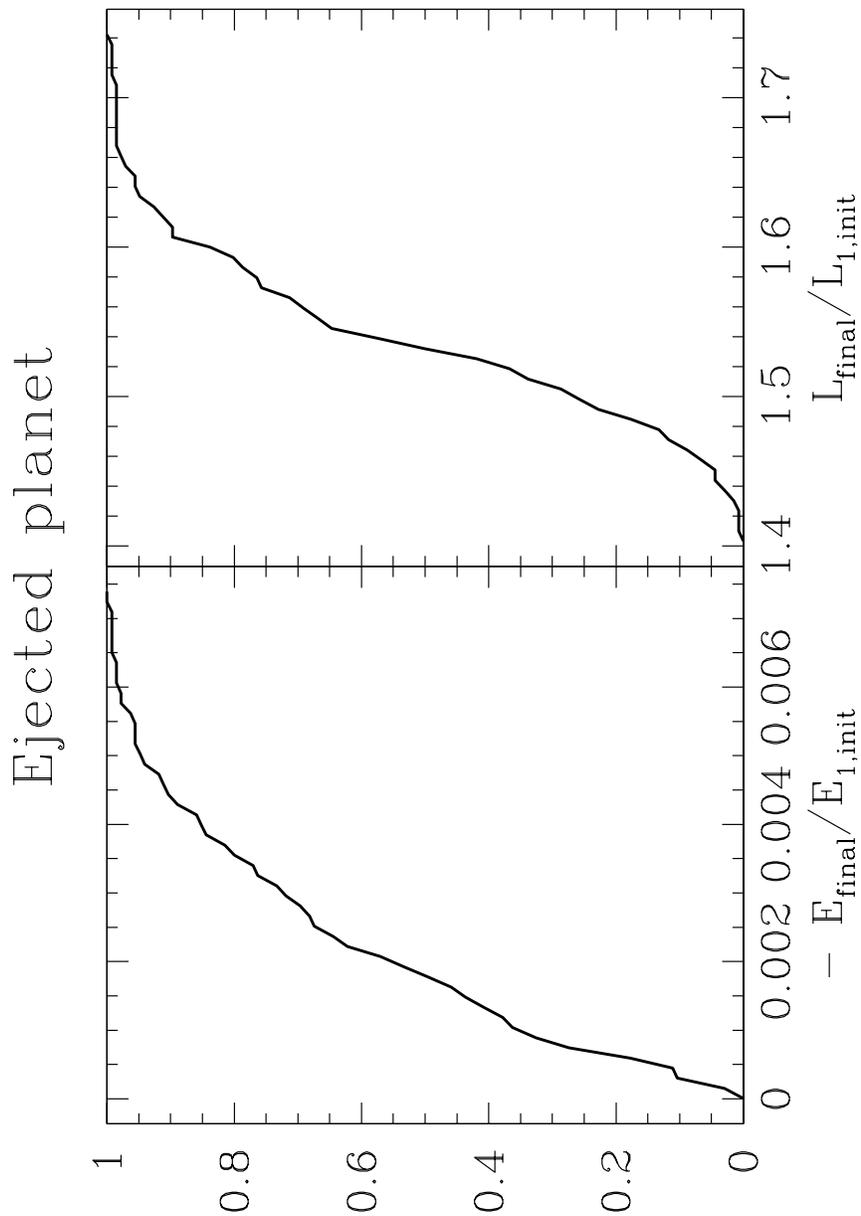} 
\figcaption{Cumulative distributions of the energy and angular momentum 
of the escaping planet in units of their initial values.  While the ejected 
planet carries very little (positive) energy, it removes a significant amount of
angular momentum from the system. \label{FigEjectEL} }
\end{figure}

We now turn to those cases where one planet is ejected from the system.
The distributions of orbital elements
for the remaining planet are shown in Fig.~\ref{FigEject}
(we do not distinguish between ejecting what was initially the inner 
vs outer planet; the outer planet is ejected most often). 
The escaping planet typically leaves the system with a very small
(positive) energy, and the final semimajor axis of the remaining
planet is therefore set by energy conservation at a value very near
$a_{\rm final}\simeq a_1a_2/(a_1+a_2)\simeq 0.56$ in our units
(See Appendix). However, the escaping
planet does carry away significant angular momentum
(Fig.~\ref{FigEjectEL}). As a result, the
distribution of final eccentricities is much broader: about 90\% of the
remaining planets have eccentricities in the range $e_{\rm
final}\simeq 0.4 - 0.8$, with a median value around 0.6. Some planets
develop very large eccentricities.  Of
great potential importance are the few percent of systems with very
small pericenter distances ($r_{\rm p,final}\la 0.1$), which may later
become tidally circularized, especially if the central star is still
on the pre-main-sequence.  Inclinations of remaining planets following
an ejection generally remain small, with about 90\% of the orbits
having $i_{\rm final}< 10^{\circ}$.

\label{StarGrazers}

In two of our numerical integrations ($\ll 1\%$) one planet 
came extremely close to the central star.  These systems were not included in
Figs.~\ref{FigTime}--\ref{FigEject}, since the
numerical integrations did not conserve energy
and angular momentum to the required precision.  However a more
careful analysis of these systems reveals that the errors accumulated
once their orbit already took them very close to the star (at pericenter
distances $r_p\la0.05$). Depending on the initial separation and the 
radius of the star, these systems could be affected by strong tidal
forces that are not included in our simulations.  In particular,
the orbits could circularize at a very small semimajor axis
$a\simeq 2r_p$, or the inner planet could be tidally disrupted
by the star.

\label{FigUnresText}

\begin{figure}
\epsscale{0.80}
\plotone{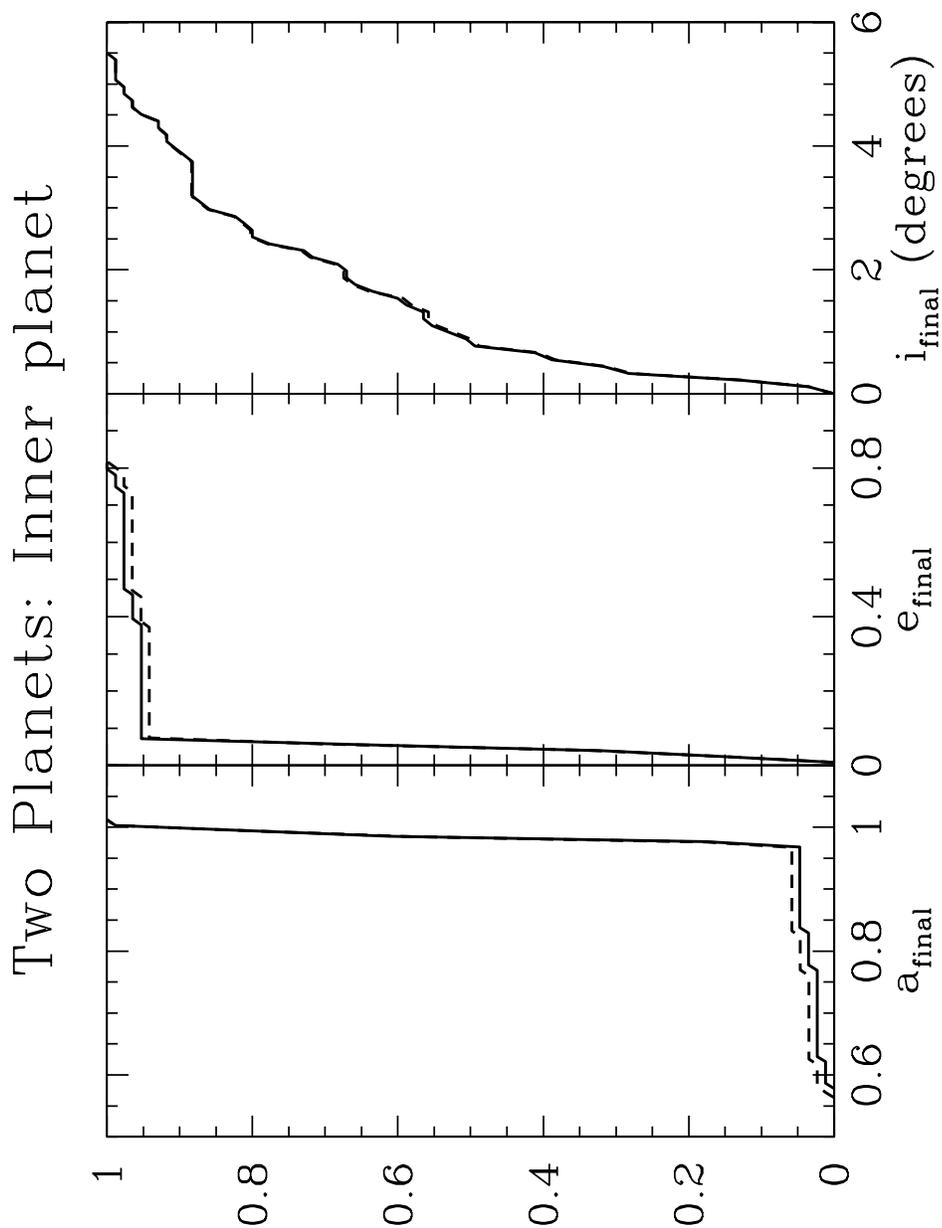}
\figcaption{Cumulative distributions of the final pericenter separation,
eccentricity, and inclination of the inner planet in systems that have
retained two bound planets by the end of the numerical integration.
The solid and dashed lines are for different planetary radii, with
$(R/a_1)\times(5\,{\rm AU}/R_{\rm J}) = 5$ and~1, respectively.\label{FigUnres1}}
\end{figure}

\begin{figure}
\epsscale{0.80}
\plotone{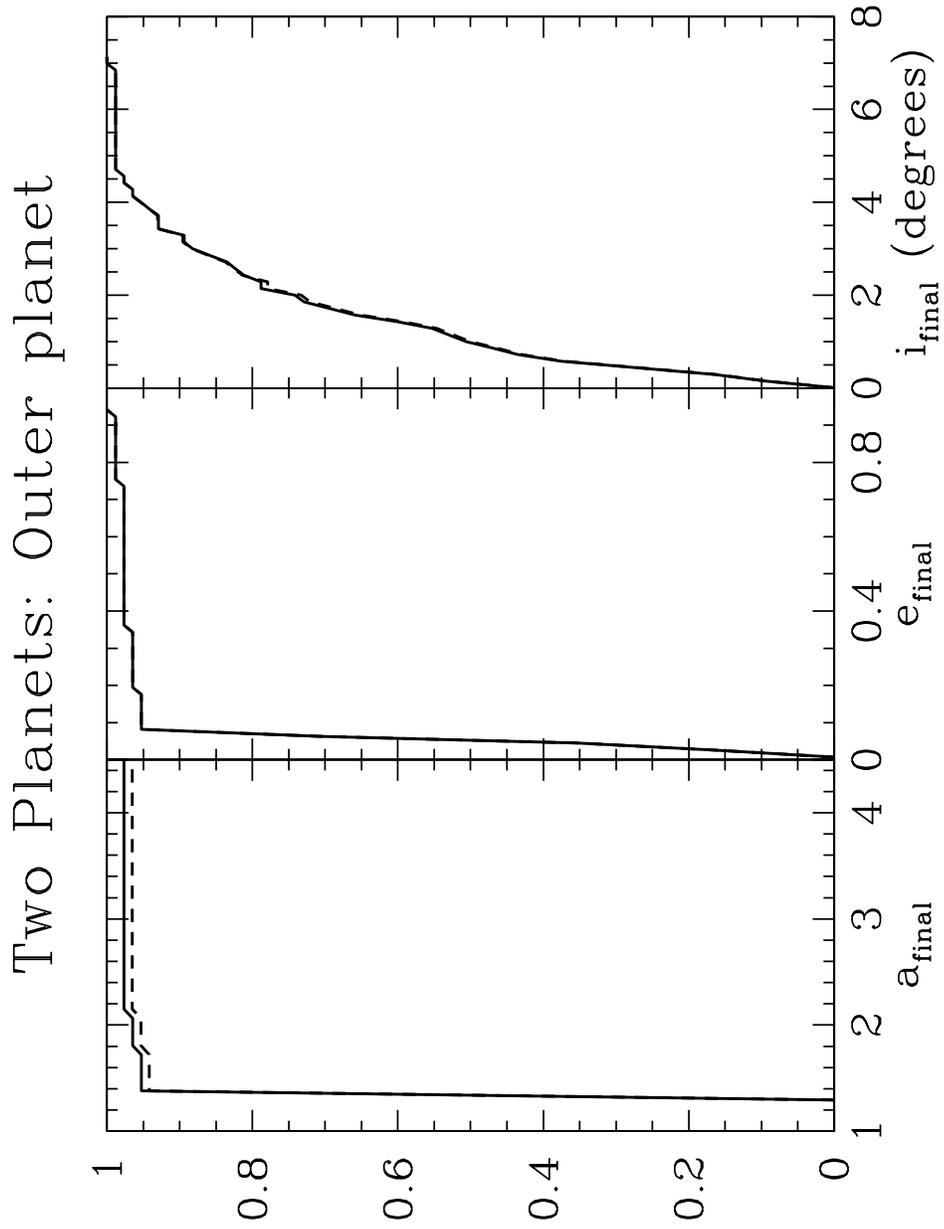} 
\figcaption{Same as Fig.~\ref{FigUnres1} but for the outer planet.\label{FigUnres2}}
\end{figure}

\begin{figure}
\plotone{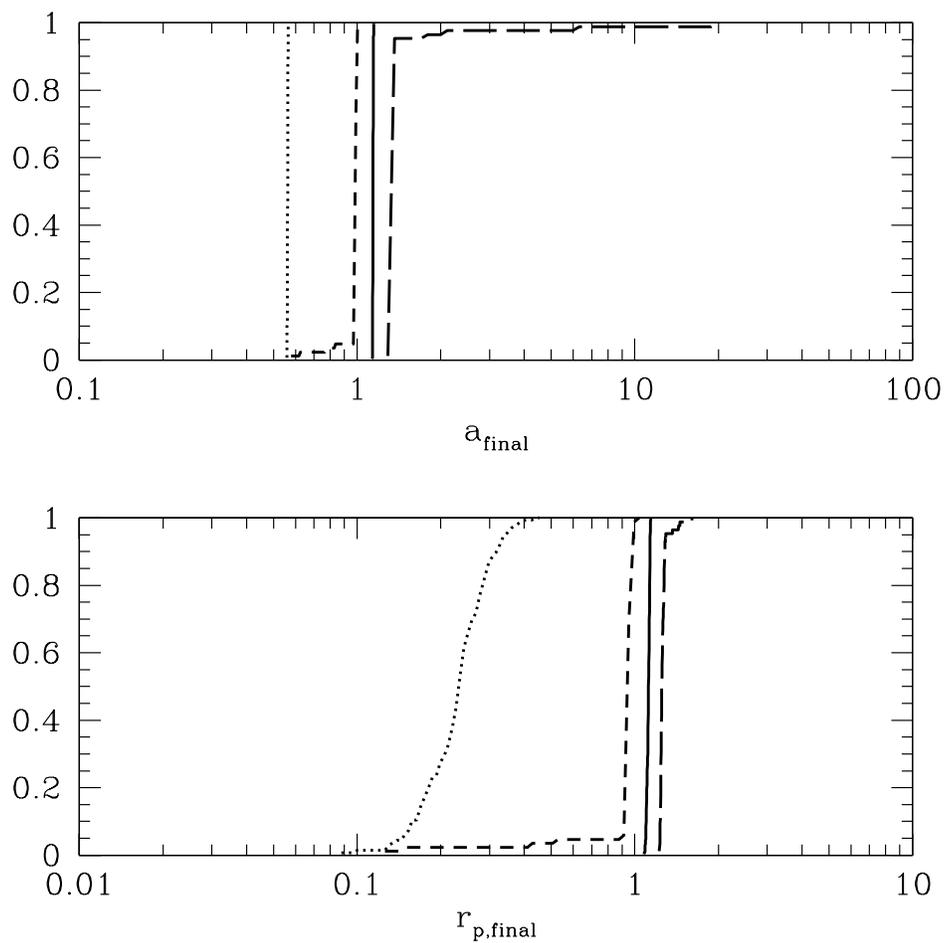}
\figcaption{Final distributions of semimajor axes and pericenter distances
compared for different outcomes.  The solid line is for collisions, 
the dotted line for ejections, and the two dashed lines for systems with both 
planets remaining.  When only one planet remains, conservation of energy 
dictates a narrow range of semimajor axes. Among systems with two planets remaining, 
a small fraction may still be on their way to dissociation, while most are in 
a quasi-stable state.\label{FigAandRp}}
\end{figure}

\begin{figure}
\plotone{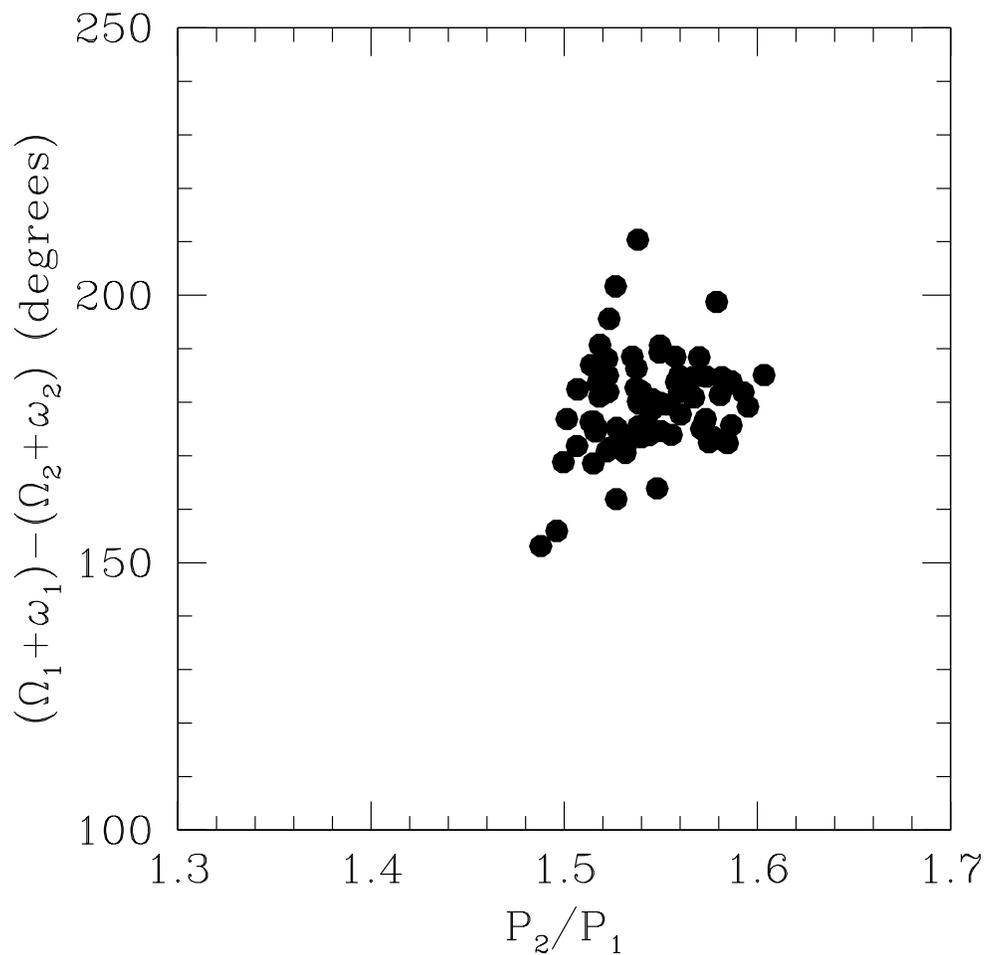}
\figcaption{Angle between pericenters vs ratio of orbital periods for the
two planets when they have remained in a stable, closely coupled 
configuration until
the end of the numerical integration (here $\omega_1$ and $\omega_2$
are the longitudes of pericenters with respect to the ascending nodes,
while $\Omega_1$ and $\Omega_2$ are the longitudes of the ascending
nodes; subscripts 1 and~2 refer to the inner and outer planets,
respectively).  The orbits are clearly locked in a nonlinear 3:2 resonance.
\label{FigUnresperi}}
\end{figure}

Finally, we study the properties of systems that are still containing
two bound planets at the end of the numerical integrations.  The
distributions of orbital elements for the inner and outer planets (as
determined at the end of the integration), are shown in
Fig.~\ref{FigUnres1} and Fig.~\ref{FigUnres2}, respectively.  These
systems can be clearly divided into two categories (see
also Fig.~\ref{FigAandRp}). About $10\%$ have a large ratio
of semimajor axes $a_{\rm outer}/a_{\rm inner}\ga 3$ and are either in a
stable hierarchical triple configuration, or on their way to the
ejection of the outer planet on a timescale exceeding the length of
our integrations (recall that we stop integrating after a time $t_{\rm
max}=2\times10^7\,{\rm yr}\,[P_1/12\,{\rm yr}]$). Note that the
secular evolution of hierarchical triple systems can take place on
extremely long timescales that are difficult to probe with direct
numerical integrations of the orbital dynamics (see, e.g., Ford {\it
et al.\/} 2000).  Most systems, however ($90\%$), retain a ratio
of semimajor axes very close to the initial value.  These systems also
retain their very small initial eccentricities and inclinations. They
clearly represent the dynamically stable region of our initial
parameter space. Closer inspection of their properties reveal that
they are in fact locked in a nonlinear resonant configuration (see,
e.g., Peale 1976) with a near 3:2 ratio of orbital periods, and
pericenters that remain anti-aligned at all times
(Fig.~\ref{FigUnresperi}). We will not discuss these systems further
in this paper, since their evolution will depend crucially on the dissipation 
processes that are still at work when the resonant configuration is
formed.

\section{Comparison with Observations}

The known extrasolar planets (Fig.~1) can be roughly divided into two groups:
those with short-period, nearly circular orbits ($a\la 0.07\,$AU) and those 
with wider and more eccentric orbits ($a\ga 0.07\,$AU).  

Many of the short-period planets, like their prototype 51~Peg, are
so close to their parent star that tidal dissipation would have likely
circularized their orbits, even if they were originally eccentric (Rasio {\it et
al.\/}~1996).  Thus, their small observed eccentricities do not provide a 
good indicator of their dynamical history. Circularization of extremely 
eccentric orbits produced
by dynamical instabilities (as originally proposed by Rasio and Ford 1996)
seems unlikely to be the dominant mechanism for producing these systems. 
Indeed, the observed frequency
of 51~Peg type systems appears much higher than would be predicted by 
such a dynamical scenario: in the observed sample the
frequency is $\sim20\%$, while
among all stars searched for planetary-mass companions it is $\sim1\%$ 
(Marcy and Butler 2000).
In contrast, on the basis of our simulations for two planets,
we would estimate that at most a few in $\sim10^3$ systems affected by
dynamical instabilities would produce an orbit eccentric enough to
be circularized by tidal dissipation at $r\la 0.07\,$AU. We note, however,
that tidal interaction with a (much larger) pre-main-sequence star, or
dissipation in a gaseous disk, could circularize orbits at a considerably
larger distance from the star, increasing the predicted frequency of
circularized systems in our scenario (see Fig.~\ref{FigAandRp}).
Observational support for the existence of a circularization mechanism
operating at distances as large as $r\sim0.2\,$AU is provided by
some of the wider systems with nearly circular
orbits, such as $\rho$ CrB (with $a\simeq 0.23\,$AU and $e<0.07$; 
see Noyes {\it et al.\/} 1997). These 
orbits are clearly too wide to have been circularized by tidal dissipation
in the star or in the planet, according to the standard theory 
(Ford {\it et al.\/} 1999). 
We note also that the observation of a single giant planet on a nearly circular orbit
does {\it not\/} imply that the parent planetary system must have been 
dynamically stable,
since a frequent outcome of a dynamical instability is a collision between
two planets, which leaves a more massive single planet on a nearly circular
orbit (assuming the initial orbits of the two planets were nearly circular;
see Fig.~\ref{FigCollide}).

\begin{figure}
\epsscale{0.90}
\plotone{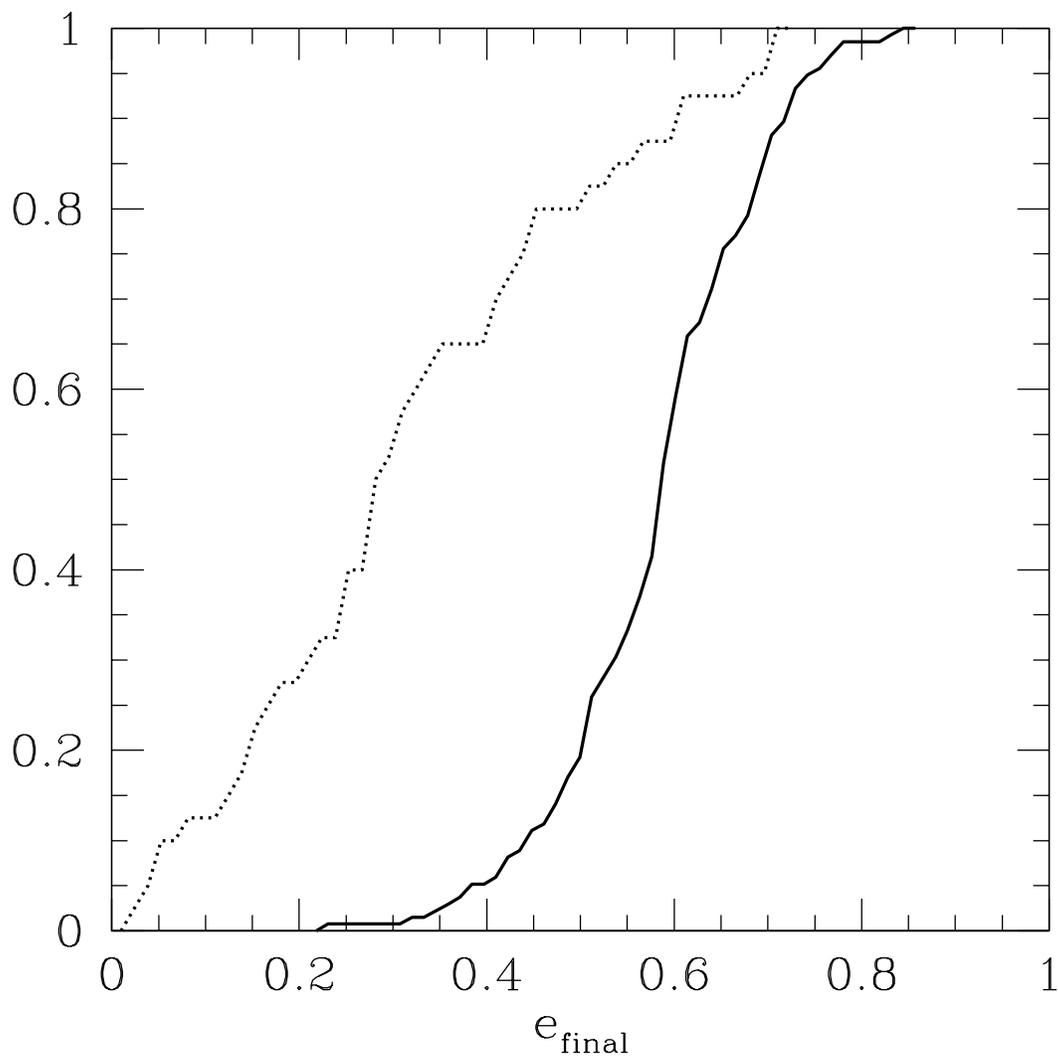} 
\figcaption{Cumulative distribution of orbital eccentricities for
the observed systems (dotted line; we included all orbits with $a>0.07\,$AU
in Fig.~1), compared to the eccentricity distribution
predicted by our numerical simulations for the remaining planet following an
ejection (solid line, as in Fig.~\ref{FigEject}). The median eccentricity of
the observed sample is about 0.3, compared to 0.6 from the simulations.
\label{Figecum}}
\end{figure}

The large eccentricities of most planets with longer periods also
require an explanation.  A planet that would have formed from a protoplanetary
disk in the standard manner is unlikely
to have developed such a large eccentricity, since dissipation in the disk
tends to circularize orbits. Dynamical instabilities leading to the
ejection of one planet while retaining another planet of comparable mass
can naturally explain the observed distribution of eccentricities.
A direct comparison between the observations and our results (Fig.~\ref{Figecum})
suggests that dynamical instabilities would actually tend to {\it overproduce\/}
highly eccentric orbits. However, since our simulations were done for two
{\it equal-mass\/} planets, they provide an {\it upper limit\/} to the actual 
distribution.
Indeed, for slightly unequal masses, the dynamical interactions will tend to 
eject preferentially the less massive planet, thereby allowing the more massive
planet to retain a higher angular momentum. We can see easily that only a small
departure from the equal-mass case
would be necessary to bring our predicted eccentricity
distribution in closer agreement with the observed one. The median eccentricity
in the observed sample is about 0.3, while it is about 0.6 in our simulations.
To reduce the eccentricity from 0.6 to 0.3 for the retained planet corresponds
to an increase in orbital angular momentum by a factor 
$(1-0.3^2)^{1/2}/(1-0.6^2)^{1/2}\simeq 1.2$. Thus a $\sim20\%$ reduction in
the angular momentum removed by the escaping planet, which could be achieved
by a $\sim 20\%$ reduction in its mass, would be sufficient to bring our
results in close agreement with the observed distribution.

Our scenario also imposes tight constraints on the distribution of
semimajor axes that are far more difficult to reconcile with observations.
What is perhaps most striking about the orbital parameters of the observed
systems with $a\ga 0.07\,$AU in Fig.~1 is the paucity of circular orbits.
However, in our scenario, we cannot avoid a certain fraction of systems
that still contain a giant planet on a nearly circular orbit, following
a collision between the two initial planets. From Fig.~\ref{FigRadius}
we see that, to avoid a significant fraction of collisions, we must
have $(R/R_{\rm J})/(a_1/5\,{\rm AU}) \la 1$. Since $a_f\simeq 0.56 a_1$
for the retained planet following an ejection, we deduce that, if most of
the observed planets on wide eccentric orbits had been retained following
the ejection of another planet (and with most avoiding a collision), their
semimajor axes should satisfy $a_f \ga 2.5\,{\rm AU}\,(R/R_{\rm J})$.
Instead, in the range of semimajor axes observed for eccentric systems, 
$a\simeq 0.07 - 3\,$AU, we would expect that collisions would be about
$\sim3$ times more frequent than ejections, implying that the fraction
of highly eccentric orbits could not exceed $\sim1/4$ of the observed
systems. Reducing the planetary radius to $R\simeq 0.1\,R_{\rm J}$ would
provide very good agreement with observations (and collisions would then 
explain nicely the existence of systems like $\rho$ CrB), but this seems 
rather implausible: even a giant terrestrial (rocky) planet with 
$m\ga 0.5\,M_{\rm J}$ would have $R \simeq 0.3\,R_{\rm J}$ 
(Guillot {\it et al.\/} 1996). Moreover, we note that the observations
of transits indicate that HD 209458b must be a hydrogen-rich gas giant
(Burrows {\it et al.\/} 2000).

\acknowledgments
We are grateful to Scott Tremaine for valuable conversations. We also thank
the referees, S.~Ida and P.~Wiegert, for many useful comments on the
original manuscript. 
This work was supported in part by NSF Grant AST-9618116 and NASA ATP Grant
NAG5-8460.  F.A.R.\ was supported in part by an Alfred P.\ Sloan Research
Fellowship.  Our computations were supported by the National
Computational Science Alliance under Grant AST980014N and utilized 
the SGI/Cray Origin2000 supercomputers at Boston University and NCSA,
and the Condor system at the University of Wisconsin.

\section*{Appendix: Simple Analytic Estimates}

Here we set $G=1$, but we retain factors of $a_1=1$ and $M=1$ (in our
units) in all equations for clarity.

\subsection*{Collisions}

If we could entirely neglect the change in orbital energy following
a collision of two planets, under our assumptions that collisions
conserve both mass and momentum, the final semimajor axis for the
new planet of mass $2m$ would be given by
\begin{equation}
a_f = -\frac{2mM}{2E_f} \simeq -\frac{mM}{(E_1+E_2)}\simeq\frac{2a_1a_2}{a_1+a_2},
\end{equation}
where $a_1$ and $a_2$ are the initial semimajor axes of the two planets
($E_f$ is the final orbital energy, $E_1$ and $E_2$ the initial orbital
energies of the two planets, and we have neglected the interaction
energy, which represents a fractional error $\sim m/M \sim 10^{-3}$).
The resulting range of $a_f/a_1$ (taking into account our small range of
initial values for $a_2/a_1$) is $\simeq 1.12-1.13$.
This is slightly {\it lower\/} than the actual range obtained from 
our simulations, where $a_f/a_1\simeq 1.13-1.15$ (see Fig.~\ref{FigCollide}),
indicating that the total orbital energy of the system {\it increases\/}
by about $1-2\%$ following a collision.

We can easily understand this result by considering the following simple 
model for a collision, suggested by our numerical results 
(see Fig.~\ref{FigExCollide}). As long as they are well outside each
other's sphere of influence (where the mutual gravitational attraction
of the two planets becomes dominant over the central star), the relative
velocity between the two planets remains always very small compared to
the escape speed from their surface. Indeed, the relative velocity is
$v_r\sim 0.5(M/a^3)^{1/2}\,\Delta a$ for two planets separated by
$\Delta a = a_2-a_1$, while the escape speed $v_e=(2m/R)^{1/2}$, giving
\begin{equation}
\frac{v_r}{v_e} \sim 3\times10^{-2} 
                \left(\frac{M/m}{10^3}\right)^{1/2}
                \left(\frac{R/a}{10^{-4}}\right)^{1/2}
                \left(\frac{\Delta a/a}{0.3}\right).
\end{equation}
The radius $r_i$ of the sphere of influence is determined by
setting $m^2/r_i^2 \sim Mm/a^2$, giving $r_i\sim (m/M)^{1/2}\,a \gg R$.
In the center-of-mass frame of the two planets, the collision
resembles a head-on collision between two planets of mass $m$
starting from a distance $r_i$ at rest.
Neglecting $(v_r/v_e)^2 \ll 1$, we see that the orbital energy change
following the collision (which leaves a single planet of mass $2m$ at
rest in the center-of-mass frame) is equal to the gravitational
binding energy $m^2/r_i$. The fractional increase in the total
orbital energy following a collision should therefore be
$\Delta E/E\sim (m^2/r_i)/(2Mm/a) \sim 0.5\,(m/M)^{1/2}$, which
is $\sim 1.5\%$ for $m/M=10^{-3}$, in close agreement with the
numerical results. Note that this argument is completely independent
of the details of the collision itself, which converts a much larger
amount of kinetic energy into heat through shocks, and a much larger
amount of gravitational binding energy of the two planets just before
impact into binding energy of the collision product.

Having determined the final semimajor axis following a collision,
we can now also estimate the final eccentricity from
conservation of angular momentum. With obvious notations we write,
for two nearly circular and coplanar initial orbits,
\begin{eqnarray}
L_f & = & \frac{2mM}{M+2m}\sqrt{(M+2m)a_f(1-e_f^2)} \\
    & = & L_1+L_2 \\
    & = & \frac{mM}{M+m}\left(\sqrt{(M+m)a_1}+\sqrt{(M+m)a_2}\right),
\end{eqnarray}
and solving for the final eccentricity gives
\begin{eqnarray}
1-e_f^2 = 
 \frac{(M+2m)\left(\sqrt{a_1}+\sqrt{a_2}\right)^2}{4a_f(M+m)}.
\end{eqnarray}
The {\it maximum\/} final eccentricity is obtained by minimizing the RHS.
For $m/M=10^{-3}$, $a_1=1$, $a_2=1.3$, and $a_f=1.15$ (the maximum
value of $a_f$, taking into account the slight increase in orbital
energy estimated above; see Fig.~8), we obtain
$e_f < 0.05$, in perfect agreement with our numerical results
(see Fig.~\ref{FigCollide}).

\subsection*{Ejections}

Similarly we can try to predict the orbital properties of the remaining
planet following an ejection. Since the ejected planet leaves the system
on a very nearly-parabolic orbit, we can estimate the final semimajor axis of the
retained planet from energy conservation,
\begin{equation}
a_f = -\frac{mM}{2E_f} \simeq -\frac{mM}{2(E_1+E_2)}\simeq\frac{a_1a_2}{a_1+a_2},
\end{equation}
using the same notations and assumptions as above.
With $a_1=1$ and $a_2=1.3$ we obtain $a_f\simeq 0.565$, which is precisely
the upper limit of the range of values, $a_f\simeq 0.558 - 0.565$, obtained
from our simulations (Fig.~9). Thus the (positive) energy carried away by the
escaping planet is at most $\simeq0.7\%$ of its initial binding energy (in
agreement with the distribution of escaping energies shown in Fig.~10).

We can again try to 
estimate the final eccentricity using conservation of angular momentum.
With the same notations as before and with $r_{pe}$ denoting the pericenter
distance of the ejected planet's parabolic orbit we have
\begin{eqnarray}
L_f & = & \frac{mM}{M+m}\left(\sqrt{(M+m)a_f(1-e_f^2)} + \sqrt{2(M+m)r_{pe}}\right) \\
    & = & L_1+L_2 \\
    & = & \frac{mM}{M+m}\left(\sqrt{(M+m)a_1}+\sqrt{(M+m)a_2}\right).
\end{eqnarray}
Solving for the final eccentricity and using eq.~(7) gives
\begin{eqnarray}
1-e_f^2 \simeq
 \frac{a_1+a_2}{a_1a_2}\,\left(\sqrt{a_1}+\sqrt{a_2}-\sqrt{2r_{pe}}\right)^2.
\end{eqnarray}
Unfortunately no simple argument can be used to predict precise
values of $r_{pe}$. Clearly, however, we expect the pericenter distance
$r_{pe}$ of the ejected planet
to be just slightly larger than the apocenter distance of the inner
(retained) planet, i.e., $r_{pe}\ga 1$. From the range of values of the final
eccentricity $e_f\simeq 0.4 - 0.8$ observed in our simulations (Fig.~9),
we deduce that $r_{pe}\simeq 1 - 1.4$.

\end{document}


In this section we wish to predict what the final orbital elements for
ejections and collisions should be, using analytic methods, and
compare them to the actual results of our simulations.

\section{Notation and basic formulas}

In the following sections we shall use notation as follows:

$a$ - semimajor axis
$e$ - eccentricity
$E$ - energy
$L$ - angular momentum
$m$ - mass of either planet
$M$ - mass of the star

The subscripts $i$ and $f$ will stand for initial and final
parameters, respectively. Subscripts $1$ and $2$ will stand for the
parameters referring to the initially inner and outer planet, in that
order.

The basic formula for energy and angular momentum of an orbit is then

\begin{equation}
E = -\frac{m}/{2a}
\:\:\:\:
L = \frac{Mm}{M+m}\sqrt{G(M+m)a(1-e^2)}
\end{equation}

\section{Collisions}

Assuming a complete conservation of energy, we can estimate the final
semimajor axis of the planet resulting from a collision:

\begin{equation}
a_f = -\frac{2m}{2E_f} = -\frac{m}{(E_1+E_2)}=\frac{2a_1a_2}{a_1+a_2}
\end{equation}

The resulting range of $a_f$ (taking into account our distribution of
initial orbital elements) is approx. $1.123-1.13$, which is slightly
lower than the actual range resulting from our simulation. This is a
consequence of our energy conservation assumption, which is not
very accurate for an inelastic collision. We can easily estimate the
actual amount of dissipated energy:

\begin{equation}
\frac{E_{actual}}{E_1+E_2} = \frac{a_{f,calculated}}{a_{f,actual}}
\end{equation}

The resulting dissipation is between $1$ and $2 \%$.

We can also estimate the final eccentricity,
assuming conservation of angular momentum in the z direction (which,
given the small inclinations, is approximately equal to its total
magnitude):

\begin{eqnarray}
L_f & = & L_1 + L_2 \\
\frac{2mM}{M+2m}\sqrt{(M+2m)a_f(1-e_f^2)} & = &
\frac{mM}{M+m}\left(\sqrt{(M+m)a_1(1-e_1^2)}+\sqrt{(M+m)a_2(1-e_2^2)}\right)
\end{eqnarray}

which gives us the formula

\begin{eqnarray}
e_f & = &
\sqrt{1-\frac{(M+2m)\left(\sqrt{a_1(1-e_1^2)}+\sqrt{a_2(1-e_2^2)}\right)^2}{4a_f(M+m)}}\\
\end{eqnarray}

This means that $e_f$ is in the range of approximately
$\sqrt{1-\frac{1.126}{a_f}}$ to $\sqrt{1-\frac{1.147}{a_f}}$.

Since the expression under the square root has to be non-negative, we
can use this equation to estimate the semimajor axis
(assuming the final eccentricity to be 0), to be in the range of
$1.126 - 1.147$, which
matches the data a little better; more than $95\%$ of the outcomes are
under $1.147$. The lower bound of $1.126$ is slightly lower than the
result of our simulations, which has a lower bound of about
$1.135$. We can use this to estimate the dissipation of energy again,
getting a result of about $1\%$.

EXPLANATION OF THIS? I HAVE NO IDEA.

\section{Ejections}

Ejections are a little harder to predict exactly, since we cannot know
how much energy the escaping planet is carryint away. Assuming it
carries none, we can get the highest possible final semimajor axis of
the remaining planet:

\begin{equation}
a_f = -\frac{m}{2E_f} = -\frac{m}{2(E_1+E_2)}=\frac{a_1a_2}{a_1+a_2}
\end{equation}

which gives the maximum of $0.565$, consistently with our simulation results.

From conservation of angular momentum, we can get the following:

\begin{equation}
e_f = \sqrt{1-\frac{L_f^2(1+m)}{a_fm^2}}=\sqrt{1-\frac{L_f^2(1+m)(a_1+a_2)}{a_1a_2m^2}}
\end{equation}

Setting the final eccentricity to be 0, we can get the maximal angular
momentum left to the bound planet to be about 0.75 in the units of
$L_{1,init}$, the initial angular momentum of the inner planet. That
sets the angular momentum carried away by the escaping planet to be in
the range of 1.38-2.14 $L_{1,init}$